\documentclass{sf2a-conf2017}
\usepackage{graphicx}
\usepackage{hyperref}
\usepackage[]{natbib}  
\usepackage{epstopdf}

\def\BibTeX{{\rm B\kern-.05em{\sc i\kern-.025em b}\kern-.08em
    T\kern-.1667em\lower.7ex\hbox{E}\kern-.125emX}}
\bibpunct{(}{)}{;}{a}{}{,}  

\newcommand{\gbp}{{G_\mathrm{BP}}}
\newcommand{\grp}{{G_\mathrm{RP}}}
\newcommand{\grvs}{{G_\mathrm{RVS}}}

\begin{document}

\TitreGlobal{SF2A 2017}


\title{Gaia: on the road to DR2}

\runningtitle{Gaia: on the road to DR2}

\author{D. Katz}\address{GEPI, Observatoire de Paris, PSL Research University, CNRS, Place Jules Janssen, F-92195 Meudon Cedex, France}

\author{A.G.A. Brown}\address{Leiden Observatory, Leiden University, PO Box 9513, 2300 RA, Leiden, the Netherlands}




\setcounter{page}{237}


\maketitle


\begin{abstract}
The second Gaia data release (DR2) is scheduled for April 2018. While Gaia DR1 had increased the number of stars with parallaxes by a factor 20 with respect to the Hipparcos catalogue, Gaia DR2 will bring another factor 500 increase, with parallaxes (and proper motions) for more than a billion stars. In addition, Gaia DR2 will deliver improved accuracy and precision for the astrometric and photometric data, $G$, $\gbp$, $\grp$ magnitudes, radial velocities, identification and characterisation of variable stars and asteroids as well as stellar parameters for stars down to $G = 17$~mag. On behalf of the teams of the Gaia-DPAC consortium, these proceedings give a foretaste of Gaia DR2, 6 months before the release.
\end{abstract}



\section{Introduction}

Gaia was launched from Kourou on a Soyouz-Fregat rocket on 19$^{th}$ December 2013. After a 4 week journey, Gaia was inserted on its nominal orbit around the second Lagrange point (L2) of the Sun-Earth system, on 14$^{th}$ January 2014. There followed a 6 month commissioning phase and finally the start of the nominal mission phase on 14$^{th}$ July 2014. For almost 4 years, Gaia has been continuously scanning the sky along 2 lines of sight simultaneously. Its two telescopes both feed 3 instruments: an astrometric instrument, a spectro-photometer and a spectrograph. The spectro-photometer uses two prisms to measure the spectral energy distribution (SED) of the sources in the blue  (BP: 330-680~nm) and in the red (RP: 640-1050~nm). The spectrograph, the {\it Radial Velocity Spectrometer} (RVS), is a medium resolving power (R$\sim$11~500) near infrared (845-872~nm) integral field spectrograph. The two fields of view and the 3 instruments are imaged on a single focal plane, made of 106 CCD detectors and almost a billion pixels. \citet{GaiaPrusti2016} provides a full description of the Gaia payload, mission and science case.\\

The development and operation of the ground-segment is under the responsibility of the Gaia Data Processing and Analysis Consortium (DPAC). The 450 members of the consortium are in charge of the calibration of the instruments, of the extraction of the astrophysical information and of the publication of the releases. The first Gaia data release (DR1), published a year ago on 14$^{th}$ September 2016 \citep{GaiaBrown2016, Arenou2017} contained parallaxes and proper motions for about 2 millions stars, positions and mean $G$ magnitudes for 1.1 billion sources and $G$-band light curves for about 3000 Cepheids and RR-Lyrae. Over the past 12 months, about 300 papers have already made use of the Gaia DR1. They address a large variety of science topics: e.g. open clusters \citep{GaiaVanLeeuwen2017}, Cepheids and RR-Lyrae \citep{GaiaClementini2017}, Magellanic cloud kinematic \citep{vanDerMarelSahlmann2016} and structure \citep{BelokurovErkalDeason2017}, dynamical influence of the bar \citep{Monari2017} and spiral arms \citep{Hunt2017}, stellar rotation \citep{Davenport2017} and many many others.\\

Coming only 1.5 years after Gaia DR1, Gaia DR2 is scheduled for April 2018. At the moment of writing these proceedings, the processing of the data has been completed and the validation is actively on-going. These proceedings present a preview of the Gaia DR2 content, 6 months before the release. Since, we are still in the validation phase, there might be some differences between this preview and the actual contents of the April release.\\

The second Gaia data release is the result of the collective and dedicated work of the 450 DPAC members.

\section{DR2 overview}
For Gaia DR2, the 22 first months of data from the nominal mission have been processed (it was 14 months for Gaia DR1). The longer time baseline did allow for a stand alone (Gaia only) astrometric solution. From Gaia DR1 to Gaia DR2, the data processing pipelines have continued their developments and optimization, with specific attention for the instruments calibrations, leading to reduced systematics and improved precisions. The cross-match, i.e. the association of a group of observations with a given source, was also upgraded. As a consequence, the Gaia DR2 source identifiers will be independent from the Gaia DR1 ones\footnote{\url{https://www.cosmos.esa.int/web/gaia/news_20170203}}. A means of tracing their evolution will be provided. The pipelines have not only been upgraded, but new functionalities have also been added ($\gbp$ and $\grp$ magnitudes from the photometric pipeline) and new pipelines activated: solar-system objects, spectroscopy and stellar parameters pipelines.

\section{Astrometry and Photometry}
For Gaia DR1, 11.2 months of data were processed. This was not enough to disentangle the parallaxes from the proper motions. In order to break this degeneracy, in Gaia DR1, a combined Tycho-Gaia Astrometric Solution (TGAS) was solved for \citep{Michalik2015, Lindegren2016}, allowing to derive the 5 astrometric parameters (positions, parallax, proper motions) for 2 million stars in common with the Hipparcos and Tycho-2 catalogues \citep{HipparcosTycho1997, vanLeeuwen2007, Hog2000}. With Gaia DR2, the longer baseline is enough to separate parallaxes from proper motions and to solve the full 5 parameter astrometric solution, using only Gaia data, for about 1 billion stars down to G=20~mag. Positions will be published for another half a billion stars. From Gaia DR1 to Gaia DR2, a lot of work has focused on improving the calibrations and in particular in the addition of colour-dependent astrometric terms as well as improved modelling of the satellite attitude and removal of the attitude disturbance (e.g. micro-meteoroids hits). The result is a reduction of the systematics and an improvement of the precision, e.g. the preliminary values for the formal errors on the parallaxes are: 30 $\mu$as at G=15, 150 $\mu$as at G=18 and 700 $\mu$as at G=20.\\

The number of stars with G magnitudes will increase from Gaia DR1 to Gaia DR2, from 1.1 billion \citep{Carrasco2016, vanLeeuwen2017, Evans2017} to about 1.5 billion. An important novelty of Gaia DR2 are the $\gbp$ and $\grp$ magnitudes which have also been measured for 1.5 billions stars. For the bright stars the expected precision is of the order of the milli-magnitude.

\section{Spectroscopy}
\begin{figure}[ht!]
\centering
\includegraphics[width=0.83\textwidth,clip]{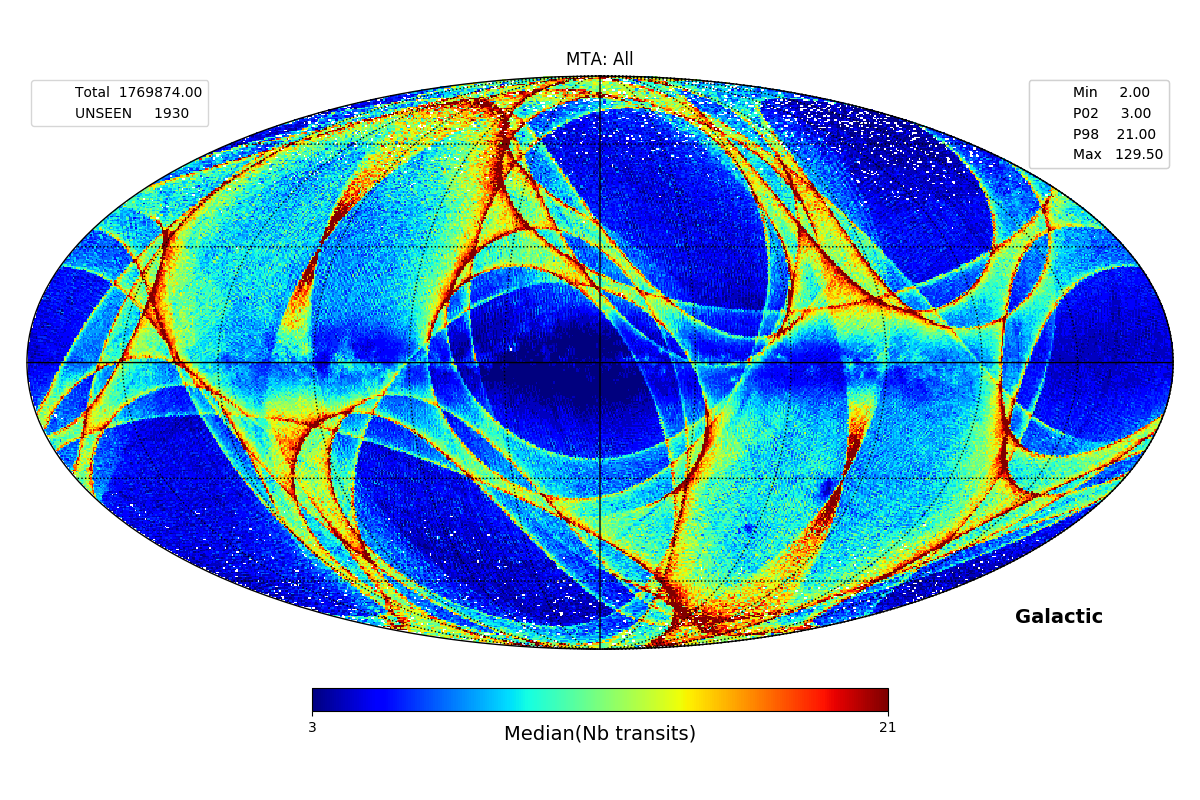}      
\caption{Distribution on the sky (Galactic coordinates) of the median (per pixel) number of RVS transits per source. The area of the pixel is $\sim$0.2~deg$^2$.}
\label{katz:fig1}
\end{figure}

\begin{figure}[ht!]
\centering
\includegraphics[width=0.83\textwidth,clip]{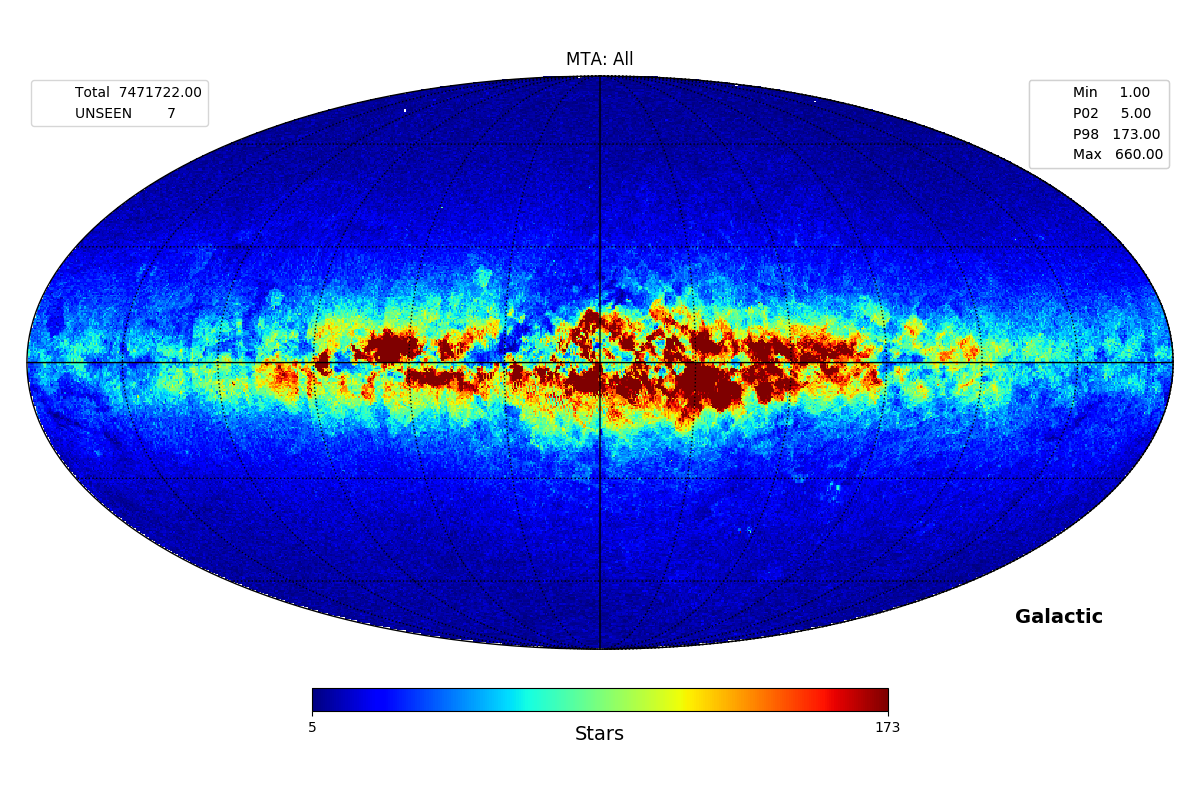}      
\caption{Stellar density map (Galactic coordinates) of the stars for which a median velocity has been derived and which have passed the validation tests (so far). The area of the pixel is $\sim$0.2~deg$^2$.}
\label{katz:fig2}
\end{figure}

Gaia DR2 will be the first Gaia release with radial velocities. For Gaia DR2, 235 millions of RVS spectra of stars brighter than $\grvs = 12$ were processed. The RVS records 3 spectra per observation (hereafter referred to as transits). The spectroscopic pipeline derives one radial velocity per transit. Therefore, about 78 millions transit radial velocities were obtained. The number of RVS transits per source is driven primarily by the Gaia scanning law and shows large excursions on the celestial sphere (figure~\ref{katz:fig1}) around the median which, in Gaia DR2 and for the RVS, is 7 transits per source. For Gaia DR2 the transit radial velocities will be combined to publish the median radial velocities of the 5 to 7 million stars fulfilling the following criteria: at least 2 RVS transits, T$_{eff}$ in the range $]3500, 7000[$~K, no large dispersion of the transit radial velocities, good data quality indicators. Figure~\ref{katz:fig2} shows the distribution on the sky of the stars which have passed the validation tests so far (2 months before the end of the validation phase).\\
 
\begin{figure}[ht!]
\centering
\includegraphics[width=0.9\textwidth,clip]{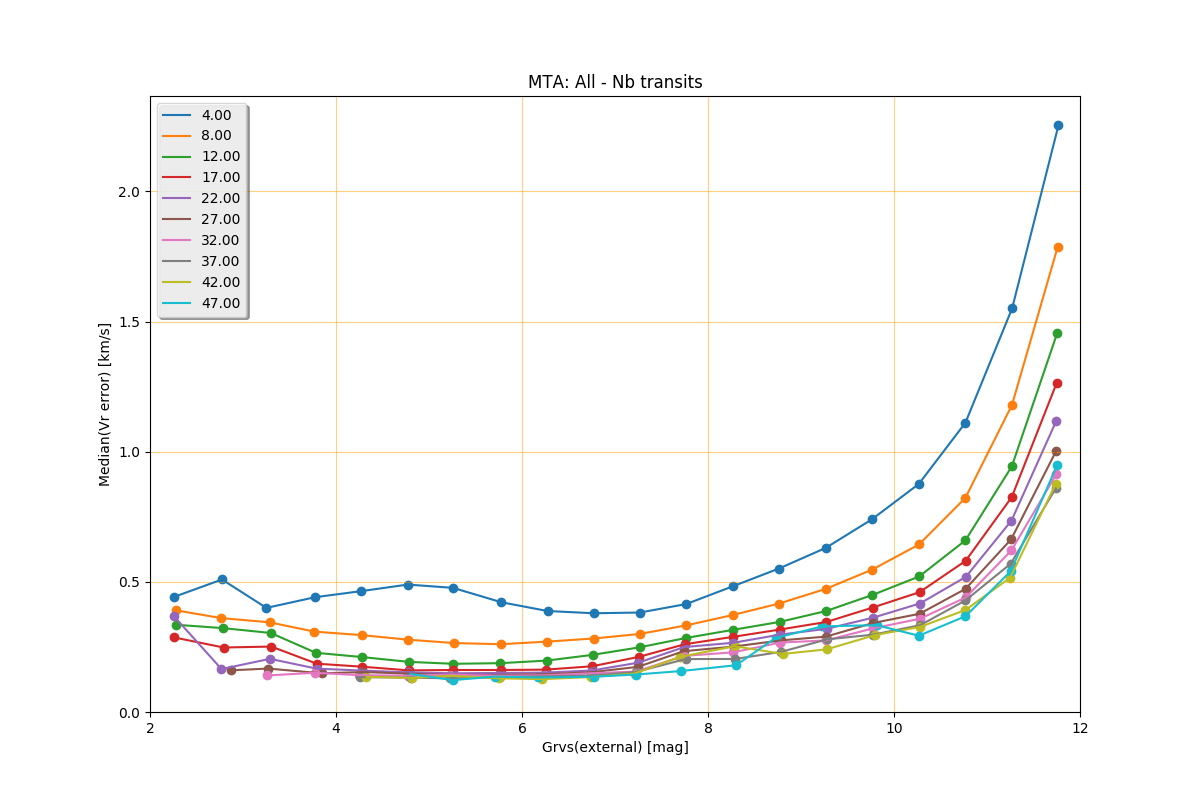}      
\caption{Estimated radial velocity precision versus $\grvs$ magnitude and for different numbers of transits.}
\label{katz:fig3}
\end{figure}

Figure~\ref{katz:fig3} shows the estimated radial velocity precision versus $\grvs$ magnitude and for different numbers of transits. On the bright side, the precision reaches a level of about 150 m/s. This is the current level of the wavelength calibration floor. It is about 6 times more precise than the pre-launch expectation of 1~km/s. This achievement was possible in particular because the RVS spectrograph (build by Astrium, now Airbus-DS) is a very stable instrument and Gaia has shown very smooth and stable behaviour during most of the mission. At the faint end the precision is not governed by the calibration residuals, but mostly by the photon noise. The precision therefore improves with the number of transits, from about 2.3~km/s at 4 transits to 0.9~km/s at 40 transits (and $\grvs=11.75$~mag).\\

\section{Asteroids, Variable stars, Stellar parameters}
Another novelty of Gaia DR2 is the publication of asteroids. Several thousands of them with more than 9 transits have been detected and characterised in the 22 months of data. They include in particular: Near-Earth Objects, Main-Belt Asteroids and Trojans.\\

During the 2 first months of the nominal mission, Gaia followed a specific scanning law, passing through the ecliptic poles at each rotation of the satellite. The {\it Ecliptic Pole Scanning Law} (EPSL) yielded a large number of transits for the stars in the vicinity of the ecliptic poles. This, and the very hard work of the Gaia-DPAC variability group, resulted in the publication of a first catalogue of Gaia variable stars in Gaia DR1, ahead of the pre-launch schedule \citep{Clementini2017, Eyer2017, GaiaClementini2017}. It contained 599 Cepheids and 2595 RR-Lyrae grouped in 38 deg$^2$ and belonging, for a large fraction, to the Large Magellanic Cloud. In Gaia DR2, the longer time baseline and therefore the larger number of transits, will enable the extension of the variability detection and characterisation pipeline to hundreds of thousands of variables over the whole sky.\\

The last novelty of Gaia DR2 is the publication of astrophysical parameters for stars. The $G$, $\gbp$, $\grp$ magnitudes and the parallaxes have been used to derive the effective temperature (T$_{eff}$), absorption (A$_V$), luminosity (L) and radius (R) of stars down to  $G=17$~mag.

\section{... and beyond Gaia DR2}
Two releases are already planned after Gaia DR2: \url{https://www.cosmos.esa.int/web/gaia/release}\\

The third Gaia data release (DR3) is scheduled mid/late 2020. In addition to the performance improvement, DR3 should also contain: source classification, stellar parameters derived from BP and RP spectral energy distributions and RVS spectra, radial velocities for $\sim$40-50 millions stars brighter than $\grvs = 14$~mag, non-single stars as well as extended catalogues of solar system objects and variable stars.\\

The fourth Gaia data release (DR4) is planned end 2022. It should reach the specified end-of-mission performance, contain all planned deliveries, including the exo-planets catalogue and the transit data. DR4 will be the final release for the nominal mission.\\

The end of the nominal mission is 2019, but Gaia current micro-propulsion fuel supplies would allow the satellite to operate until 2024. A proposal for a 5 year extension of the mission has therefore been made. If the extension is accepted, one (or a few) catalogue(s) will follow DR4, delivering performances beyond the original mission goals.

\section{Conclusions}
Gaia DR2 represents a big step forward: positions, parallaxes and proper motions for a billion stars, G, $\gbp$, $\grp$ for 1.5 billions stars, median radial velocities for 5 to 7 million stars as well as asteroids, variable stars and stellar parameters.\\

Gaia DR1 and soon Gaia DR2 are the result of more than ten years of collective work by the DPAC consortium members for designing, implementing, optimizing, testing and operating the Gaia ground-based pipelines and validating the data produced.\\

\bibliographystyle{aa}  
\bibliography{katz} 

\end{document}